\documentclass[preprint,pteplogo]{ptephy_v2}

\preprintnumber{2025-0214} 
\usepackage{hyperref}

\usepackage{doi}

\begin{document}

\title{Improvement to Generalized Separable Expansion Method in Lippmann-Schwinger Equation}


\author{Hiroyuki Kamada$^{a,b}$}
\affil{$^a$Depatment of Physics, Faculty of Engineering, Kyushu Institute of Technology, Kitakyushu 804-8550, Japan, \\
$^b$Research Center for Nuclear Physics, Osaka University,\\
10-1 Mihogaoka, Ibaraki, Osaka 567-0047, Japan \email{kamada@rcnp.osaka-u.ac.jp} }





\begin{abstract}%
Realistic nucleon-nucleon (NN) potentials are generally not in separable form, but there is a way to convert them into separable potentials, called the generalized separable expansion (GSE). When the separable potential is substituted into a three-body Faddeev equation, which generally has two Jacobi momenta, the integral equation is conveniently reduced to a one-variable integral equation. The two-body scattering t-matrix of the conventional GSE does not have an exact singularity at the energy threshold of the two-body bound state. The newly introduced GSE improves this by treating the singularity analytically.
\end{abstract}
\subjectindex{xxxx, xxx}

\maketitle

\section{Introduction}
To investigate the wave function and the scattering matrix in a nucleon-nucleon (NN) system, one must solve the Sch\"odinger equation or the Lippmann-Schwinger equation. The latter equation is often treated in momentum representation, and becomes a Fredholm integral equation. Kowalski-Noyes showed \cite{Noyes1965,Kowalski1965} a method to directly solve this equation while keeping it as continuous variables. Some ingenuity is required to reduce the continuous variables when dealing with few-body systems.

In general, when a potential is represented in a separable form \cite{Yamaguchi1954}, the degree of freedom to handle continuous variables is reduced, and so an integral equation of $n$ dimensions generally becomes an equation of $(n-1)$ dimensions, which accelerates the calculation speed.
In addition to this convenience, there is an advantage that if the analyticity of the separable form factor is improved, the singularity and cut problems of complex integral equations can be solved by changing real variables into complex variables.
As a typical example, the three-body problem of a three-nucleon system has been calculated using the so-called Faddeev equation \cite{Faddeev1961}, which generally has two Jacobi momenta in the center mass system. 
After partial wave expansion, the integral equation included with separable potential, is conveniently reduced to a one-variable integral equation (Amado-Lovelace equation)\cite{Amado1963, Lovelace1964}.

With the help of supercomputers, efforts have been made to solve three-body scattering problems without approximations \cite{Gloeckle}, but when moving on to four-body problems \cite{Kamada2001,Uzu2003} or even many-body problems, the separable expansion method is a good guide for serious calculations. Methods for converting a realistic potential into a separable potential include the EST expansion method \cite{EST}, the USE method \cite{USE}, and the Generalized Separable Expansion (GSE)  \cite{Oryu1974} which is the subject of this paper. In general, the accuracy of the expansion method improves as its rank increases. A test of the accuracy of the GSE method has been performed \cite{GSEkentei} using the Reid Soft Core potential \cite{RSC}.
The potential after separable expansion becomes an approximate potential as the rank increases, gradually approaching the original potential.

In this paper, we address the problem that the conventional GSE expansion method does not precisely reproduce the bound state even at rank 1, and we will show the solution of this problem.
In the next section, we introduce an improved GSE. In Section 3, we analytically show that the scattering matrix has a singular point of order 1 at the energy of the bound state, and also prove that the wave function of the bound state calculated from the original potential is exactly given from the improved GSE potential.
Our conclusion and outlook are presented in Section 4.

\section{An Improved Generalized Separable Expansion}

The bound state (deuteron) $\psi_d$ satisfies the following Schr\"odinger equation;
\begin{eqnarray}
    (E_d - {p^2 \over m} ) ~\psi_d(\vec p)= \int V(\vec p, \vec p~') ~ \psi_d (\vec p~') { d\vec p~' \over (2\pi)^3} 
    \label{eq1}
\end{eqnarray}
where $E_d$ ($<0$) is a binding energy of the bound state and $V(\vec p, \vec p~') $ is a NN potential. $m$ is nucleon mass and $\vec p$ is the relative momentum between two nucleons.   After a partial-wave decomposition, we represent Eq. (\ref{eq1}) as
\begin{eqnarray}
    (E_d - {p^2\over m}) \psi_d (p) = \int _0^\infty V(p,p') \psi_d (p')  {p'^2d p' \over 2 \pi^2}
\end{eqnarray}
Let us define $g_0(p)$ as a function;

\begin{eqnarray}
    g_0(p)\equiv (E_d-{p^2\over m}) \psi_d(p)
\end{eqnarray}
And $\tilde V  (p,p') $ is also defined as
\begin{eqnarray}
    &&  \tilde V(p,p') = V(p,p') -g_0(p){1 \over C} g_0(p') 
    \label{eq_4}
\end{eqnarray}
with 
\begin{eqnarray}
    && C\equiv \int _0^\infty g_0(p) {1\over E_d -{p^2\over m} } g_0(p) {p^2 d p \over 2 \pi^2 } \cr 
    && = \int _0 ^\infty \psi_d (p) (E_d -{p^2 \over m} ) \psi_d (p) {p^2 d p \over 2\pi^2}  \cr &&
    = \int_0^\infty  \int _0 ^\infty \psi_d(p) V(p,p') \psi_d(p') {p^2 dp \over 2\pi^2} {p'^2 dp' \over 2\pi^2}=\langle V \rangle.
\end{eqnarray}
This $C$ means the potential expectation of the bound state. 
We perform the conventional GSE expansion \cite{Oryu1974} of $\tilde V$ and add the separable term $g_0(p){1\over C}g_0(p')$.
\begin{eqnarray}
 V_{GSE} (p,p') && = \sum _{i=1}^N \sum _{j=1}^N \tilde V(p,k_i) \tilde \Lambda_{i,j} \tilde  V(k_j,p')+g_0(p){1\over C}g_0(p')\cr &&
    = \sum _{i=0}^N \sum _{j=0}^N \tilde V(p,k_i) \Lambda_{i,j} \tilde  V(k_j,p')= \sum_{i=0}^N\sum_{j=0}^N g_i(p) ~\Lambda_{i,j}~g_j(p'),
\label{eq_GSE}
\end{eqnarray}
where the momenta $k_i$ are the so-called Bateman parameters \cite{Bateman1921}. The Bateman parameters $k_i$ must be chosen appropriately so that the parameters are not too close to each other. The rank of the matrix $\Lambda_{i,j}$ is $N+1$. The matrix $\tilde \Lambda_{i,j}$  (rank $N$) is given by the matrix inversion of $\tilde V (k_j,k_i)$.
\begin{equation}
\Lambda_{i,j} \equiv 
\left\{ 
  \begin{alignedat}{4}   
 & {1\over C} & ~~~~~~ i=0,~j=0 \\
  &0                    & ~~~~~~ i>0,~j=0 \\
  &0                    &~~~~~~~i=0,~j>0 \\
 & \tilde \Lambda _{i,j}\equiv [\tilde V (k_j,k_i)]^{-1}  &~~~~~~~  i>0,~j>0 
  \end{alignedat} 
  \right.
\end{equation}
and 
\begin{equation}
g_i (p)\equiv 
\left\{ 
  \begin{alignedat}{2}   
 & g_0(p) & ~~~~~~ i=0 \\ 
 &\tilde V(k_i,p) =\tilde V(p,k_i) & ~~~~~~~  i>0
  \end{alignedat} 
  \right. 
\end{equation}
The improved GSE potential $V_{GSE}$ in Eq. (\ref{eq_GSE}) inherits the properties of the original potential $V$;
\begin{eqnarray}
  &&   V_{GSE} (p,k_i) = \tilde V(p,k_i) + g_0(p)~{1\over C} ~g_0(k_i)
  =V(p,k_i) , \cr 
   &&  V_{GSE} (k_i,p') = \tilde V(k_i,p') + g_0(k_i)~{1\over C} ~g_0(p')
  =V(k_i,p') .
\end{eqnarray}
Of course, since the Lippmann-Schwinger equation (LS) of the energy half shell is satisfied, the phase shift can be calculated exactly.
With the scattering matrix $t_{GSE}$, the following equation can be shown;
\begin{eqnarray}
    && t_{GSE}(k_i,p';E_i )= V_{GSE} (k_i, p')+\int _0^\infty  V_{GSE}(k_i,p'') {1\over E_i - {{p''}^2 \over m}  +i\epsilon} ~ t_{GSE}(p'',p';E_i) {{p''}^2d p''\over 2\pi^2} \cr
    && =V(k_i, p')+\int _0^\infty V(k_i,p'') {1\over E_i - {{p''}^2 \over m}  +i\epsilon} ~t_{GSE}(p'',p';E_i) {{p''}^2d p''\over 2\pi^2}
\label{GSE_LS}
\end{eqnarray}
with
\begin{eqnarray}
    E_i \equiv {k_i^2 \over m}.
\end{eqnarray}
Comparing Eq.(\ref{GSE_LS}) to the original scattering t-matrix $t$ which  satisfies with the following the LS equation, 
\begin{eqnarray}
     t(k_i,p';E_i )=V(k_i, p')+\int _0^\infty V(k_i,p'') {1\over E_i - {{p''}^2 \over m} +i\epsilon} ~t(p'',p';E_i) {{p''}^2d p''\over 2\pi^2}.
\label{LS}
\end{eqnarray}
The relation between $t_{GSE}$ and $t$ is expressed by the following equations.
\begin{eqnarray}
    t_{GSE}(k_i,p';E_i)=t(k_i,p';E_i)
\end{eqnarray}
and 
\begin{eqnarray}
      t_{GSE}(k_i,k_i;E_i)=t(k_i,k_i;E_i).
\end{eqnarray}
In general, the two are not equal for the momentum values $p$ other than $k_i$, but the approximation can be improved by increasing the rank N sufficiently.
\begin{eqnarray}
    t_{GSE}(p,p';E_i) \ne t(p,p';E_i) ~~~~ p\ne k_i, ~~{\rm nor}~~ p'\ne k_j  
\end{eqnarray}

\section{Singularity of t-matrix}

The GSE scattering matrix (t-matrix) $t_{GSE} (E)$ is given the separable form as
\begin{eqnarray}
    t_{GSE}(p,p';E)= \sum_{i=0}^N \sum_{j=0}^N g_i(p)~\tau_{i,j}(E)~g_j(p),
    \end{eqnarray}
    where $E$ is an arbitrary c.m. energy between 2 nucleons.
The improved GSE potential $V_{GSE} $ in Eq. (\ref{eq_GSE}) is substituted into the LS equation in Eq. (\ref{GSE_LS}) for arbitral energy $E$. We have 
\begin{eqnarray}
    \tau_{i,j}(E) = \Lambda_{i,j} + \sum_{k=0}^N \sum _{l=0}^N~\Lambda_{i,k} ~I_{k,l}(E) ~ \tau_{l,j} (E),
\end{eqnarray}
and sandwich by the inverse matrices $[\Lambda]^{-1}$ and $ [\tau]^{-1}$, namely,
\begin{eqnarray}
    \Lambda^{-1}_{i,j} = \tau^{-1}_{i,j}(E) + I_{i,j}(E)
    \label{tauinv2}
\end{eqnarray}
with 
\begin{eqnarray}
    I_{i,j}(E)\equiv  \int_0^\infty g_i(p){1\over E-{p^2\over m}+i\epsilon} g_j(p) {p^2dp \over 2\pi^2 }
\label{Iij}
\end{eqnarray}

The integral $I_{i,j}(E) $ in Eq.(\ref{Iij}) is calculated in case of $j=0$ ( see Appendix \ref{apdx1});
\begin{eqnarray}
    \tau^{-1}_{i,0}(E) = \left\{ 
  \begin{alignedat}{2}   
 & (E-E_d+i\epsilon) \int_0^\infty \psi_d^2(p) {E_d -{p^2\over m} \over E- {p^2 \over m}+i\epsilon } {p^2 dp \over 2\pi^2} & ~~~~~~ i=0 \\ 
 &(E-E_d+i\epsilon ) \int_0^\infty  {\tilde V(k_i,p) \psi_d(p)  \over E-{p^2\over m}+i\epsilon }  {p^2 dp \over 2\pi^2} & ~~~~~~~  i>0
  \end{alignedat} 
  \right. 
  \label{tauinv}
\end{eqnarray}
From this it is clear that $\tau_{i,j}(E)$ is proportional to $1\over E-E_d+i\epsilon $ and diverges at $E=E_d$.
\begin{eqnarray}
    \tau_{i,j} \propto {1\over E-E_d +i\epsilon}.
\end{eqnarray}

The original bound wave function $\psi_d$ satisfies the following LS equation in $E=E_d$ (see Appendix \ref{apdx2}). 
\begin{eqnarray}
   &&  {1\over E_d-{p^2\over m} }  \int_0^\infty  V_{GSE} (p,p') ~\psi_d(p')  {p'^2 dp' \over 2\pi^2} =\psi_d(p) 
   \label{eq_wf}
\end{eqnarray}

In other words, the bound state $\psi_{GSE}$ calculated from $V_{GSE}$ is equal to the original bound state $\psi_d$.
\begin{eqnarray}
    \psi_{GSE}(p)=\psi_d (p).
\end{eqnarray}

\section{Conclusion and Outlook}


The difference from conventional GSE \cite{Oryu1974} is that the bound state is treated specially. In other words, the bound state $\psi_d$ is used to prepare $\tilde V$ by removing that state from the original potential $V$ in Eq.(\ref{eq_4}). Following the traditional GSE prescription, we expand $\tilde V$ into a separable form, and add a term that generates the bound state in Eq. (\ref{eq_GSE}). It has been shown that by using the improved GSE potential, the t-matrix has a singularity of first rank at the energy point of the bound state, and the bound state can be accurately reproduced from the improved GSE potential.

In recent years, it has been converted to a low-momentum potential and applied to few-body systems \cite{Fujii2004}. It is possible to separate Hilbert spaces below and above a selected momentum \cite{Bogner2002}. It is shown that performing a USE separation expansion\cite{USE} on this low-momentum potential significantly improves the separability \cite{separability,separability2}.
Using the improved GSE potential discussed in this paper and expanding the low-momentum potential, an available expansion with high separability might be expected.

\section*{Acknowledgment}

 This work is supported by Japan Society for the Promotion of Sciene (JSPS) KAKENHI Grants No. JP22K03597.


\appendix

\section{a proof for Eq.(\ref{tauinv})}
\label{apdx1}

We start from Eq.(\ref{tauinv2}). 
\begin{eqnarray}
    \tau^{-1}_{i,j}(E)= \Lambda^{-1}_{i,j} - I_{i,j} (E)
\end{eqnarray}
In the case of $j=0$ we have ;
\begin{eqnarray}
    && \tau^{-1}_{0,0}(E)= C- \int_0^\infty g_0(p){1\over E-{p^2\over m}+i\epsilon } g_0(p) {p^2 dp \over 2 \pi^2} \cr 
    && = C- \int_0^\infty \psi_d^2(p) {(E_d-{p^2\over m})^2 \over E-{p^2\over m}+i\epsilon } {p^2 dp \over 2\pi^2} \cr 
    && =C-  \int_0^\infty \psi_d^2(p)(E_d-{p^2\over m}) {(E_d-E-i\epsilon )+( E-{p^2\over m}+i\epsilon ) \over E-{p^2\over m}+i\epsilon } {p^2 dp \over 2\pi^2} \cr
    && = C-(E_d-E-i\epsilon ) \int_0^\infty \psi_d^2(p) {( E_d-{p^2\over m} ) \over E-{p^2\over m}+i\epsilon } {p^2 dp \over 2\pi^2} - C \cr 
    && = (E-E_d+i\epsilon )\int_0^\infty \psi_d^2(p){ E_d-{p^2\over m}\over E-{p^2\over m}+i\epsilon } {p^2dp \over 2\pi^2}.
\end{eqnarray}
In the case of $j\ne 0$ we have 
\begin{eqnarray}
    &&\tau^{-1}_{0,j}(E)= 0 - \int _0^\infty g_0 (p) {1\over E-{p^2\over m}+i\epsilon} g_j(p) \cr 
    && =  -\int_0^\infty \psi_d(p) (E_d-{p^2\over m}){1 \over E- {p^2\over m}+i\epsilon} \tilde V(p,k_j)  {p^2 dp \over 2 \pi^2} \cr
    && =  -\int_0^\infty \psi_d(p) {(E_d-E -i \epsilon)+  (E-{p^2 \over m}+i \epsilon)  \over E- {p^2\over m}+i\epsilon} \tilde V(p,k_j)  {p^2 dp \over 2 \pi^2} \cr
    && = (E-E_d +i\epsilon) \int _0^\infty \psi_d(p) { 1 \over E-{p^2\over m}+i\epsilon} \tilde V (p, k_j) {p^2dp \over 2\pi^2} - \int_0^\infty \psi_d(p) \tilde V(p,k_j) {p^2 dp \over 2 \pi^2} \cr 
    && = (E-E_d +i\epsilon) \int _0^\infty \psi_d(p) { 1 \over E-{p^2\over m}+i\epsilon} \tilde V (p, k_j) {p^2dp \over 2\pi^2} -0.
\end{eqnarray}
The last equality is because of
\begin{eqnarray}
    &&\int_0^\infty \psi_d(p) \tilde V(p,k_j) {p^2 dp \over 2 \pi^2} = 
    \int _0^\infty \psi_d(p) \left( V(p,k_j) - \psi_d (p) {(E_d-{p^2\over m})(E_d -{k_j^2\over m}) \over C }\psi_d(k_j)\right) {p^2 dp \over 2 \pi^2} \cr
    && = \int _0^\infty \psi_d(p)  V(p,k_j)  {p^2 dp \over 2 \pi^2} 
    -\int _0^\infty \psi_d^2(p) {(E_d-{p^2\over m})(E_d -{k_j^2\over m}) \over C }\psi_d(k_j){p^2 dp \over 2 \pi^2} \cr
    && = (E_d -{k_j^2 \over m})\psi_d (k_j) - (E_d-{k_j^2\over m}) \psi_d(k_j) = 0.
    \label{a1}
\end{eqnarray}

\section{a proof for Eq.(\ref{eq_wf})}
\label{apdx2}

Let's start with the left side of Eq.(\ref{eq_wf}).
\begin{eqnarray}
   &&  {1\over E_d-{p^2\over m} }  \int_0^\infty V_{GSE} (p,p') ~\psi_d(p')  {p'^2 dp' \over 2\pi^2} \cr 
&& ={1\over  E_d -{p^2\over m} } \int_0^\infty \left( \sum_{i=1}^N\sum_{j=1}^N g_i(p) \tilde \Lambda_{i,j}g_j(p') +g_0(p){1\over C}g_0(p')\right) \psi_d(p') {p'^2 dp'\over 2\pi^2 } \cr
   && = {1\over  E_d -{p^2\over m} } \int_0^\infty  \sum_{i=1}^N\sum_{j=1}^N g_i(p) \tilde \Lambda_{i,j}g_j(p') \psi_d(p') {p'^2 dp'\over 2\pi^2} 
   + \psi_d(p) {1\over C} \int_0^\infty \psi_d(p') (E_d-{p'^2\over m})\psi_d(p') {p'^2 dp'\over 2\pi^2 } \cr 
   && = {1\over  E_d -{p^2\over m} } \int_0^\infty  \sum_{i=1}^N\sum_{j=1}^N \tilde V(p,k_i) \tilde \Lambda_{i,j} \tilde V(k_j,p') \psi_d(p') {p'^2 dp'\over 2\pi^2} 
   + \psi_d(p) 
\end{eqnarray}
If the first term disappears, the proof is complete, but it has already been proven in Eq. 
(\ref{a1}) that the term becomes 0. Therefore, Eq. ( \ref{eq_wf} ) is proven.

\end{document}